\documentclass[sn-nature]{sn-jnl}

\usepackage{graphicx}%
\usepackage{multirow}%
\usepackage{amsmath,amssymb,amsfonts}%
\usepackage[title]{appendix}%
\usepackage{booktabs}%

\newcommand{\OII}{\hbox{[{\rm O}\kern 0.1em{\sc ii}]}}
\newcommand{\OIII}{\hbox{[{\rm O}\kern 0.1em{\sc iii}]}}
\newcommand{\Ha}{\hbox{{\rm H}\kern 0.1em{$\alpha$}}}
\newcommand{\Hb}{\hbox{{\rm H}\kern 0.1em{$\beta$}}}
\newcommand{\Hbeq}{\hbox{{\rm \footnotesize H}\kern 0.1em{\footnotesize $\beta$}}}
\newcommand{\Hg}{\hbox{{\rm H}\kern 0.1em{$\gamma$}}}
\newcommand{\HI}{\hbox{{\rm H}\kern 0.1em{\sc i}}}
\newcommand{\HIeq}{\hbox{{\rm \tiny H}\kern 0.1em{\tiny \sc i}}}
\newcommand{\HII}{\hbox{{\rm H}\kern 0.1em{\sc ii}}}

\newcommand{\aap}{Astron. Astrophys.}   
\newcommand{\aaps}{Astron. Astrophys. Suppl.}   
\newcommand{\apj}{Astrophys. J.}   
\newcommand{\apjl}{Astrophys. J. Lett.}   
\newcommand{\apjs}{Astrophys. J. Suppl. Ser.}   
\newcommand{\araa}{Annu. Rev. Astron. Astrophys.}   
\newcommand{\mnras}{Mon. Not. R. Astron. Soc.}   
\newcommand{\nat}{Nature} 
\newcommand{\pasa}{Publ. Astron. Soc. Aust.}   
\newcommand{\pasp}{Publ. Astron. Soc. Pac.}   
\newcommand{\sci}{Science} 

\begin{document}

\title[Disk--CGM transition in emission]{An emission map of the disk--circumgalactic medium transition in starburst IRAS 08339+6517}

\author*[1,2,3]{\fnm{Nikole M.} \sur{Nielsen}}\email{nikkimnielsen@gmail.com}
\author[1,2]{\fnm{Deanne B.} \sur{Fisher}}
\author[1,2]{\fnm{Glenn G.} \sur{Kacprzak}}
\author[4]{\fnm{John} \sur{Chisholm}}
\author[5]{\fnm{D. Christopher} \sur{Martin}}
\author[1,2,6,7]{\fnm{Bronwyn} \sur{Reichardt Chu}}
\author[8]{\fnm{Karin M.} \sur{Sandstrom}}
\author[8]{\fnm{Ryan J.} \sur{Rickards Vaught}}

\affil*[1]{\orgdiv{Centre for Astrophysics and Supercomputing}, \orgname{Swinburne University of Technology}, \orgaddress{\city{Hawthorn}, \state{VIC}, \postcode{3130}, \country{Australia}}}

\affil[2]{\orgdiv{The Australian Research Council Centre of Excellence for All Sky Astrophysics in 3 Dimensions (ASTRO 3D)}, \orgaddress{\country{Australia}}}

\affil[3]{Homer L. Dodge Department of Physics and Astronomy, The University of Oklahoma, 440 W. Brooks St., Norman, OK 73019, USA}

\affil[4]{\orgdiv{Department of Astronomy}, \orgname{University of Texas at Austin}, \orgaddress{\city{Austin}, \state{TX}, \postcode{78712}, \country{USA}}}

\affil[5]{\orgdiv{Cahill Center for Astrophysics}, \orgname{California Institute of Technology}, \orgaddress{\city{Pasadena}, \state{CA}, \country{USA}}}

\affil[6]{Centre for Extragalactic Astronomy, Department of Physics, Durham University, South Road, Durham DH1 3LE, UK}

\affil[7]{Institute for Computational Cosmology, Department of Physics, Durham University, South Road, Durham DH1 3LE, UK}

\affil[8]{\orgdiv{Department of Astronomy \& Astrophysics}, \orgname{University of California, San Diego}, \orgaddress{\street{9500 Gilman Drive}, \city{La Jolla}, \state{CA}, \postcode{92093}, \country{USA}}}

\abstract{Most of a galaxy's mass is located out to hundreds of kiloparsecs beyond its stellar component. This diffuse reservoir of gas, the circumgalactic medium (CGM), acts as the interface between a galaxy and the cosmic web that connects galaxies. We present kiloparsec-scale resolution integral field spectroscopy of emission lines that trace cool ionized gas from the center of a nearby galaxy to 30~kpc into its CGM. We find a smooth surface brightness profile with a break in slope at twice the 90\% stellar radius. The gas also transitions from being photoionized by {\HII} star-forming regions in the disk to being ionized by shocks or the extragalactic UV background at larger distances. These changes represent the boundary between the interstellar medium (ISM) and the CGM, revealing how the dominant reservoir of baryonic matter directly connects to its galaxy.}


\maketitle


The vast reservoir of diffuse gas surrounding galaxies known as the circumgalactic medium (CGM) contains $\sim70\%$ of all mass that is not dark matter \citep{tumlinson_2017, somerville_2015, f-g_2023, werk_2014}. The CGM was discovered, and is almost exclusively studied, using quasar absorption line probes of galaxy outskirts. This method has provided a rough census of the gas in the CGM \citep{werk_2014}, which is crucial to reveal the constituents of the Universe. However, quasars only probe a single pencil beam sightline per galaxy and their random alignment relative to foreground galaxies means that direct observations of CGM substructure have remained elusive. One such structure is the transition between a galaxy's disk and its CGM \citep{kacprzak_2013, bland-hawthorn_2017}, where gas accreting from the cosmic web connects to the interstellar medium (ISM) to fuel future star formation and outflowing gas from stellar feedback is ejected from the disk into the surrounding CGM. Here we observe a single galaxy and its inner CGM to $\sim 0.3R_{\rm vir}$ with kiloparsec-scale resolution integral field spectroscopy of {\OII}, {\Hb}, and {\OIII} optical emission lines, which trace cool $10^4$~K gas. By directly imaging the CGM, we obtain the equivalent of thousands of quasar sightlines around a single galaxy. This controls for galaxy-to-galaxy variations that introduce scatter in the CGM properties inferred with quasar absorption lines.

\begin{figure*}[h]%
\centering
\includegraphics[width=\linewidth]{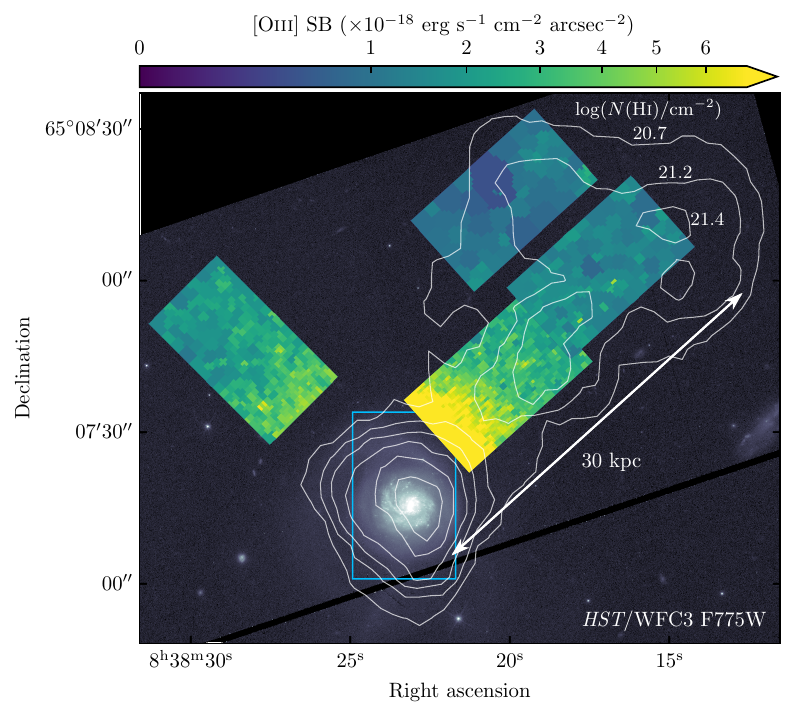}
\caption{{\bf Spatial distribution of ionized gas in the CGM at kiloparsec scales.} Emission from {\OIII}~$\lambda5007$ in the CGM of IRAS08 extends to at least $30$~kpc from the galaxy center. The blue rectangle represents the field-of-view of the KCWI pointing covering the galaxy disk (emission map not shown). {\HI} contours indicate levels of constant {\HI} column density from the VLA \citep{cannon_2004}, where a filament extends from IRAS08 towards a smaller companion galaxy $60$~kpc away.}\label{fig:map}
\end{figure*}

IRAS 08339+6517 (hereafter IRAS08) is a starbursting disk galaxy with stellar mass $\log (M_{\ast}/{\rm M_{\odot}}) \sim 10$ located at $z = 0.019$ (${d_{\rm L} \sim 83}$~Mpc). HST optical images (Fig.~\ref{fig:map}) show a face-on ($i\sim15^{\circ}-20^{\circ}$) spiral morphology that is compact for its mass \citep{mosleh_2013}, where $90\%$ of the starlight is contained within a radius of $r_{90} = 2.4$~kpc. Unlike normal spirals it has quite extreme properties (Table~\ref{tab:galprops}), with a star formation rate (${\rm SFR}=12.1$~M$_{\odot}$~yr$^{-1}$) that is $\sim10$ times higher than typical for its mass and stellar populations that are dominated by very young ($\sim4-6$~Myr) stars \citep{lopez-sanchez_2006, oti-floranes_2014, fisher_2022}. IRAS08 has a remarkably strong outflow with a velocity of $\sim400$~km~s$^{-1}$ in optical emission lines \citep{reichardt-chu_2022a} and $1,000$~km~s$^{-1}$ observed in UV absorption \citep{chisholm_2016}, where $\sim2/3$ of the wind mass flux is being driven by a kiloparsec-sized region located in a nuclear ring near the galaxy center \citep{reichardt-chu_2022a}. The galaxy also has a flat or shallow metallicity profile \citep{lopez-sanchez_2006, fisher_2022}. While we do not know for certain in IRAS08, galaxies that exhibit such metallicity profiles alongside enhanced star formation are often assumed to be acquiring significant amounts of gas \citep{peeples_2009}. Indeed, observations of the molecular gas with NOEMA \citep{fisher_2022} indicated a rapid inflow of gas to the center of the disk that is fueling the very strong starburst and subsequently strong outflows \citep{fisher_2022, peeples_2009}. VLA observations of the {\HI} gas around IRAS08 identified a filament extending out to $\sim40$~kpc from the galaxy (Fig.~\ref{fig:map}) and containing $70\%$ of the neutral gas in the system \citep{cannon_2004}. There is a nearby companion at a projected separation of $\sim60$~kpc with a mass $\sim1/10$ the mass of IRAS08 \citep{lopez-sanchez_2006}. While the interaction may be enhancing the star formation, previous work did not find evidence that it is impacting the kinematics or morphology of IRAS08 as measured by the concentration and asymmetry of the starlight \citep{fisher_2022}.

\begin{table}[h]
\caption{IRAS08 Galaxy Properties.}\label{tab:galprops}
\begin{tabular}{lclc}
\toprule%
Property & Value & Units & Reference\\
\midrule
R.A.  & 08:38:23.180 & & \\
Dec   & $+$65:07:15.20 & & \\
$z_{\rm gal}$ & 0.01911 &  &\\
Inclination, $i$ & $15-20$ & deg & \citep{reichardt-chu_2022a}\\
SFR & $12.1\pm1$ & M$_{\odot}$~yr$^{-1}$ & \citep{fisher_2022} \\
{\HI} Mass & $9.0_{-0.1}^{+0.2}$ & $\log(M_{{\HIeq}}/M_{\odot})$ & \citep{cannon_2004} \\
Stellar Mass & $10.0\pm0.1$ & $\log(M_{\ast}/M_{\odot})$ & \citep{fisher_2022} \\
Halo Mass\footnotemark[1] & $11.5\pm{0.1}$ & $\log(M_{\rm vir}/M_{\odot})$ & \\
$R_{\rm vir}$\footnotemark[2] & $115\pm4$ & kpc & \\
$r_{90}$ & 2.4 & kpc & \citep{reichardt-chu_2022a}\\
$12 + \log({\rm O/H})$ & $8.45\pm0.10$ & & \citep{lopez-sanchez_2006} \\
\botrule
\end{tabular}
\footnotetext[1]{Estimated from the stellar mass using halo abundance matching relations for redshifts $0<z<0.2$ \citep{girelli_2020}.}
\footnotetext[2]{Calculated from the halo mass using the \citep{bryan_norman_1998} relation.}
\end{table}

\begin{figure*}[h]%
\centering
\includegraphics[width=\linewidth]{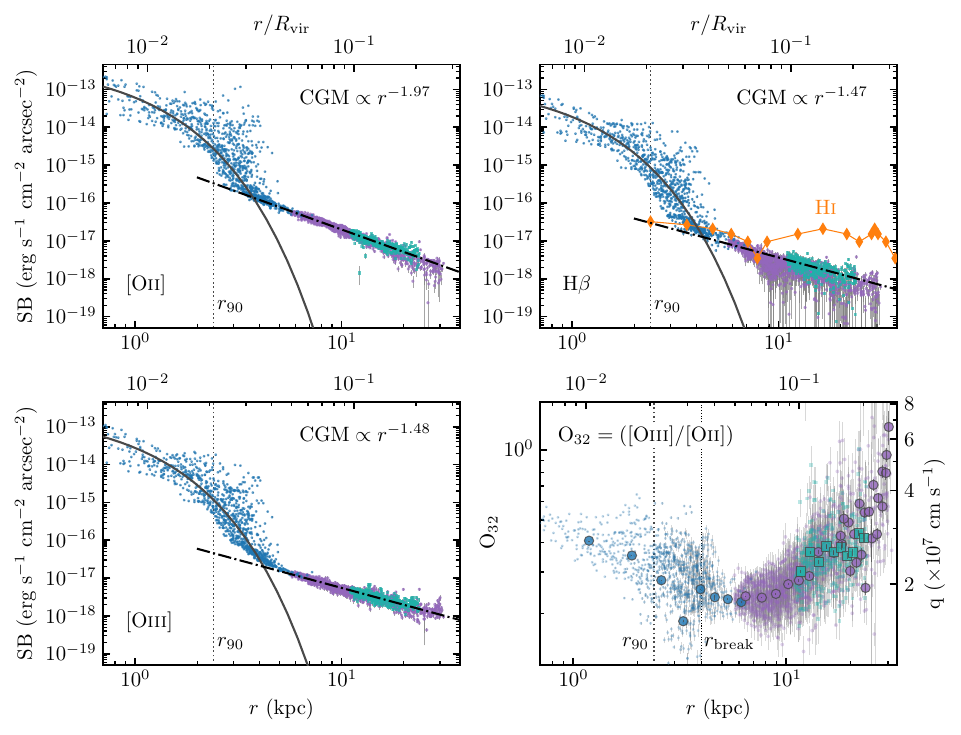}
\caption{{\bf Surface brightness and ionization radial profiles.} The galaxy (blue points) is fit by an exponential (black solid line), while the CGM is fit by a power law (black dot-dashed line) in (top left) {\OII}, (top right) {\Hb}, and (bottom left) {\OIII}. CGM emission points are colored by whether they are on (purple) or off (teal) the {\HI} filament. Error bars are $1\sigma$ uncertainties. The extended {\HI} filament column density profile (orange diamonds) \citep{cannon_2004} is arbitrarily scaled to match {\Hb} at the disk edge. (bottom right) The gas ionization indicated by ${\rm O}_{32}$ changes gradient at a radius roughly twice as far as the transition in surface brightness from the disk to the CGM ($2r_{\rm break}$). The ionization parameter, $q$, was calculated from ${\rm O}_{32}$ assuming the polynomial fit \citep{kewley_2002} for a metallicity of $\log (Z/Z_{\odot}) = 0.5$, where the IRAS08 disk has $\log (Z/Z_{\odot}) = 0.64$ and it is expected that the metallicity of the CGM is lower, decreasing with distance. Large points indicate spaxels binned by radius.}\label{fig:profiles}
\end{figure*}

We observed IRAS08 and its CGM with the Keck Cosmic Web Imager (KCWI) \citep{kcwi} in five pointings covering an area of $\sim460$~kpc$^2$ and reaching distances of $\sim30$~kpc from the galaxy both on and off the {\HI} filament (Fig.~\ref{fig:map}). We detect optical emission lines $\OII~\lambda\lambda 3727, 3729$ and $\OIII~\lambda 5007$ covering the entire area of all five pointings and $\Hb~\lambda 4861$ covering $\sim90\%$ of the area. The high covering fraction of oxygen emitting gas demonstrates that not only is the gas glowing in emission, but also that stellar feedback has been efficient at populating the CGM of IRAS08 with metals out to at least a third of the galaxy’s virial radius ($0.3R_{\rm vir}$). From Fig.~\ref{fig:profiles}, the surface brightness in all three emission lines decreases monotonically with distance from IRAS08. In contrast to the neutral hydrogen traced by the {\HI} filament, there is no significant evidence of azimuthal variation in the ionized gas (also see Fig.~\ref{fig:profile_zoom}). 

In Fig.~\ref{fig:profiles} the emission line surface brightness profile in the galaxy disk is well-fitted by an exponential with a scale radius of $0.45$~kpc in all three emission lines (Table~\ref{tab:fits}). The CGM at larger distances is well-fitted by a power law with roughly ${\rm SB} \propto r^{-2}$ in {\OII} and ${\rm SB} \propto r^{-1.5}$ in {\Hb} and {\OIII}. The transition between the galaxy exponential profile and the CGM power law occurs at $r_{\rm break} \sim 4$~kpc ($\sim 0.04R_{\rm vir}$). This corresponds to roughly twice $r_{90}$ and is consistent with the break in starlight measured in previous deep, broad-band imaging of IRAS08 \citep{lopez-sanchez_2006}. A similar power law behavior was found in the stacked CGM of SDSS galaxies with {\Ha} emission, where ${\rm SB} \propto (r/50~{\rm kpc})^{-1.9}$ out to $1.5$~Mpc \citep{zhang_2016}. Comparing to absorption lines is more difficult since absorption column densities are sensitive to low density gas ($\propto n_{\rm H}$), while emission traces the densest gas ($\propto n_{\rm H}^2$). Since the IRAS08 {\Hb} CGM emission is ${\rm SB} \propto r^{-1.47}$, this implies a volume emission density profile of ${\rm SB} \propto r^{-2.47}$. Converting to a volume gas density gives $n_{\rm H} \propto r^{-1.24}$, which is a slightly steeper relation than the volume gas density profile of $n_{\rm H} \propto r^{-0.8}$ inferred from absorption lines \citep{werk_2014}. The densest gas in the CGM appears to drop off more rapidly with radius than less dense gas.

\begin{table}[h]
\caption{Exponential and power law fit parameters to IRAS08 surface brightness profiles.}\label{tab:fits}
\begin{tabular*}{\textwidth}{@{\extracolsep\fill}cccccc}
\toprule%
 & \multicolumn{2}{c}{Galaxy Exponential\footnotemark[1]} & \multicolumn{2}{c}{CGM Power Law\footnotemark[2]} \\
\cmidrule(lr){2-3} \cmidrule(lr){4-5}%
Emission & $a_{\rm exp}$ ($\times 10^{-13}$) & $r_{\rm 0,exp}$ & $r_{\rm 0,pow}$ & $k$ & $r_{\rm break}$ \\
Line     & (erg~s$^{-1}$~cm$^{-2}$~arcsec$^{-2}$) & (kpc) & (kpc) & & (kpc [$R_{\rm vir}$]) \\
\midrule
{\OII}  & $5.7 \pm 0.5$ & $0.45 \pm 0.02$ & $4.41 \pm 0.02$ & $-1.97 \pm 0.01$ & $3.8~[0.03]$ \\
{\Hb}   & $1.7 \pm 0.1$ & $0.45 \pm 0.02$ & $1.05 \pm 0.03$ & $-1.47 \pm 0.02$ & $4.3~[0.04]$ \\
{\OIII} & $2.6 \pm 0.2$ & $0.44 \pm 0.02$ & $1.41 \pm 0.02$ & $-1.48 \pm 0.01$ & $4.3~[0.04]$ \\
\botrule
\end{tabular*}
\footnotetext[1]{${\rm SB} = a_{\rm exp}~e^{-r/r_{\rm 0,exp}}$, where $a_{\rm exp}$ is the central surface brightness and $r_{\rm 0,exp}$ is the scale radius.} 
\footnotetext[2]{${\rm SB} = (r/r_{\rm 0,pow})^k$, where $r_{\rm 0,pow}$ is the radius where ${\rm SB} = 10^{-16}$~erg~s$^{-1}$~cm$^{-2}$~arcsec$^{-2}$ and $k$ is the slope.}
\end{table}

\begin{figure*}[h]%
\centering
\includegraphics[width=\linewidth]{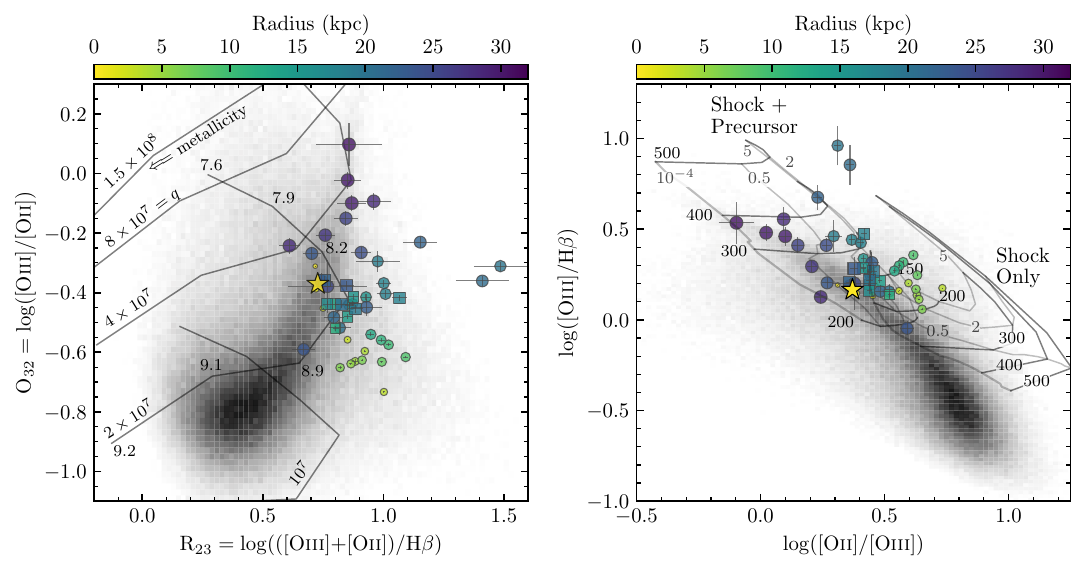}
\caption{{\bf Emission line ratios of CGM gas do not behave similar to galaxies.} Photoionization and shock diagnostic line ratios for $z=0$ SDSS galaxies \citep{sdss_dr7} (gray), the IRAS08 disk within $r_{90}$ (yellow star), and the IRAS08 CGM (colored and sized by radius). Circles are on the {\HI} filament whereas squares are off the filament and represent spaxels that have been binned by radius. Error bars are $1\sigma$ uncertainties. (left) Photoionization tracks \citep{kewley_review} (black lines) are shown for several ionization parameters ($q=10^7-1.5\times10^8$) and metallicities ($12+\log ({\rm O/H})=7.6-9.2$). (right) Shock models \citep{allen_2008} (black lines) are plotted for a range of shock velocities ($v_{\rm shock}=150-500$~km~s$^{-1}$) and magnetic parameters ($B/n^{1/2}=10^{-4}-5$~$\mu$G~cm$^{3/2}$) assuming a solar abundance and density $n=1$~cm$^{-3}$.}\label{fig:ratios}
\end{figure*}

Fig.~\ref{fig:profiles} (bottom right) shows the ratio ${\rm O}_{32} = {\OIII} / {\OII}$, which is typically used to trace the gas ionization \citep{kewley_review}, as a function of distance from IRAS08. The ${\rm O}_{32}$ ratio decays from higher ionization at the center of the galaxy to lower ionization as far as $\sim6$~kpc. This trend is typical in galaxy disks, where the number density of ionizing {\HII} regions decreases with distance from the galaxy \citep{kewley_review}. If {\HII} regions are the only source heating the gas then the ionization would continue decreasing as the surface density of ionized hydrogen decreases. However, the behavior changes beyond $\sim 6$~kpc, where the line ratio increases as a function of distance from the galaxy to at least $30$~kpc and this trend is present regardless of the {\HI} filament (Fig.~\ref{fig:profile_zoom}). The break in the ${\rm O}_{32}$ profile occurs at roughly twice the radius that the surface brightness transitions from an exponential profile to a power law. This change may indicate that an alternative mechanism is driving the emission in the CGM or, since ${\rm O}_{32}$ is also driven by metallicity, that there is a break in the gas metallicity. 

In Fig.~\ref{fig:ratios} we examine the relationship between ${\rm O}_{32}$ and ${\rm R}_{23} = (\OII + \OIII)/ \Hb$, which provides an estimate of the gas metallicity \citep{kewley_2002}. The IRAS08 disk within $r_{90}$ is consistent with SDSS galaxies and standard photoionization model tracks \citep{kewley_review}. However, gas located beyond $r_{90}$ no longer behaves like local Universe galaxies with typical radiation fields. For gas between $r_{90}$ and $\sim 6$~kpc, ${\rm O}_{32}$ decreases and ${\rm R}_{23}$ is elevated compared to the disk (the radial dependence of ${\rm R}_{23}$ is shown in Fig.~\ref{fig:profile_ratios}). This ${\rm O}_{32}$ trend then reverses direction beyond $\sim6$~kpc, where ${\rm O}_{32}$ increases and ${\rm R}_{23}$ continues to be elevated on average compared to the disk and the model tracks. The further from the galaxy the gas is located, the less it behaves like standard photoionization models. Two alternative, and plausible, mechanisms ionizing the gas are shocks and the extragalactic UV background (UVB). In the right panel of Fig.~\ref{fig:ratios} we show that shock models \citep{allen_2008} with velocities of $150-200$~km~s$^{-1}$ are consistent with the line ratios in the CGM within $6$~kpc (Fig.~\ref{fig:profile_ratios} shows the radial profile of {\OIII}/{\Hb}). This could be a result of gas accreting from the CGM onto the ISM. A few of the regions at a distance of $\sim20$~kpc have $\OIII / \Hb \sim 10$, which is somewhat large compared to typical shocks. These could instead be ionized by the UVB. While the shape of the ionizing spectrum from the UVB is uncertain, it is a much harder spectrum than thermal emission because it is likely dominated by AGNs at low redshift. For comparison, CGM gas traced by absorption lines is typically assumed to be photoionized primarily by the UVB, especially beyond $50$~kpc \citep{werk_2014, fumagalli_2011}. It is therefore likely that, in comparison to shock models and other galaxies like Makani \citep{rupke_2019, rupke_2023}, the IRAS08 CGM is shocked between $r_{90}$ and 6 kpc and ionized by the UVB or some other non-shock mechanism beyond that. 

To provide insight into the physical origin of this gas, we compare the CGM emission beyond $r_{\rm break}$ in IRAS08 to three other known environments of diffuse ionized gas in galaxies. First, the warm ionized medium (WIM) is not photoionized by {\HII} regions, similar to the IRAS08 CGM, but is only found within $\sim2$~kpc of the disk in the Milky Way \citep{gaensler_2008}. This is well within $r_{\rm break}$ for IRAS08 and thus a WIM is unlikely to dominate the CGM emission. Second, observations of outflows have been made to very large distances \citep{rupke_2019, zabl_2021}. Outflows, however, are ejected perpendicularly to galaxy disks in classic bipolar shape and they typically do not have extremely large opening angles in the local Universe \citep{shopbell_1998, westmoquette_2011, mcpherson_2023}. For a bipolar geometry in a face-on galaxy, we should detect the outflow both in front of and behind IRAS08. One side of the Makani outflow \citep{rupke_2019, rupke_2023} at similar distances is brighter and has lower mean ${\OIII}/{\Hb}$ and ${\OIII}/{\OII}$ line ratios than we observe in IRAS08. Additionally, while the outflow in IRAS08 has velocities on the order of $\sim400$~km~s$^{-1}$ \citep{reichardt-chu_2022a}, the gas near the galaxy is consistent with lower shock velocities ($150-200$~km~s$^{-1}$), which is also lower than the shock velocities found for the Makani outflow ($300-400$~km~s$^{-1}$) \citep{rupke_2023}. The face-on geometry of IRAS08 makes it difficult to fully rule out outflowing gas as the source of the emission, however, it is unlikely to dominate the CGM emission beyond $r_{\rm break}$. Finally, if the {\HI} filament is infalling, this infall extends only in the direction of the nearby companion and has a constant column density profile. However, the ionized gas shows no evidence of enhanced emission along the {\HI} filament, the emission line ratios do not correlate with the filament location, and the surface brightness decreases with distance. Our observations suggest that the ionized gas is decoupled from the neutral gas. Instead, the emission beyond $r_{\rm break}$ is most likely a diffuse CGM component that arose from an active history of gas flowing in and out of the galaxy and is not dominated by one physical mechanism. 

The ubiquitous emission around IRAS08 implies a significant cool CGM gas mass. A key property required to estimate this mass is the electron density, $n_e$. Quasar absorption line probes of the cool CGM \citep{werk_2014} and independent estimates from fast radio bursts \citep{prochaska_2019} estimate a low value of $n_e \sim 10^{-3}$~cm$^{-3}$. For our observations, only the central galaxy KCWI pointing has sufficient spectral resolution to resolve the {\OII} doublet where the line ratio can be used to measure $n_e$. At the edge of this pointing ($r = 7$~kpc) the {\OII} line ratio is below the low-density limit \citep{kewley_2019}, indicating that $n_e < 30$~cm$^{-3}$. We therefore cannot make a direct estimate of the mass, but we can use our {\Hb} luminosity profile fit and mass estimates from quasar absorption lines to place constraints on the electron density of the cool emitting CGM. To estimate a gas mass, we assume spherical symmetry in all gas properties and that the {\Hb} surface brightness power law from Fig.~\ref{fig:profiles} holds between $r_{\rm break}$ and $R_{\rm vir}$ (the rough outer boundary of the CGM). A constant $n_e = 10^{-3}$~cm$^{-3}$ and unity volume filling factor gives a total ionized gas mass three orders of magnitude larger than the estimated cool CGM mass for $L^{\ast}$ galaxies from absorption line estimates \citep{tumlinson_2017, werk_2014, stocke_2013} and, more importantly, larger even than the dark matter halo mass of IRAS08. If we instead assume $n_e \sim 0.3$~cm$^{-3}$ we infer a total ionized gas mass of $M_{R{\rm vir}} \sim 10^{11}$~M$_{\odot}$, which is similar to the absorption line estimate. The clumpiness (or volume filling factor) of the gas will further complicate this. If the clumpiness is high, then this allows for even larger electron densities for the emitting gas. Overall, our observations combined with the absorption and fast radio burst results suggest a wide range of densities in the CGM, which is also seen in recent simulations \citep{nelson_2020, dutta_density}. 

While the electron density is highly uncertain in our CGM observations, the radial light profile and, by extension, the shape of the density profile are strongly constrained in the inner CGM ($0.3R_{\rm vir}$), where the gas is likely most sensitive to the galaxy's star formation activity \citep{stern_2021}. The opposite is true for quasar absorption line surveys, which must statistically extrapolate the radial density profile from single quasar sightlines with large scatter in absorption column densities, introducing a significant source of uncertainty in the derived ionized gas mass. However, the profile shape measured here is consistent with that used in previous absorption line studies \citep{werk_2014} (although our power law is more steep) and with theoretical predictions \citep{lan_2019, nelson_2020}, reinforcing and validating the assumed geometries of these previous works. 

Our observations represent a significant step forward in obtaining a census of the largest baryonic mass reservoir of galaxies. We find empirical and quantifiable differences in the gas distribution and ionization between the ISM and the CGM in a single galaxy and make direct comparisons between the cool ionized and cold neutral gas components. We also confirm predictions from recent theory \citep{corlies_2016, piacitelli_2022} that the CGM is directly observable to scales of $\sim 50$~kpc with current ground-based instrumentation. Our observations serve as a motivator and set expectations for future UV space missions to open a new window into the main component of baryonic mass in the Universe \citep{astro2020}.

\section*{Methods}\label{sec:methods}

\subsection*{Observations}

Multi-wavelength observations of IRAS08 were obtained with WFC3 on HST (galaxy and CGM imaging), the VLA ({\HI} emission mapping) and KCWI on Keck II (galaxy and CGM IFU spectroscopy). Observations with HST/WFC3 (PID 16749) were conducted in the UVIS F775W filter totalling $2,820$~s, which traces the total star light of the galaxy. We employed a post-flash of $4$~s to mitigate for low charge transfer efficiency (CTE) and used a three-point dither pattern. The data were reduced using the standard reduction pipeline AstroDrizzle to correct for cosmic rays, CTE, and geometric distortions. An {\HI} map around IRAS08 and its companion was obtained with the VLA on 8 and 9 December 2002 in the C configuration for $370$ minutes \citep{cannon_2004}. We used WebPlotDigitizer (v4.6; https://automeris.io/WebPlotDigitizer) to extract the column density contours from \cite{cannon_2004}, which are overlaid on Fig.~\ref{fig:map}. 

The galaxy itself was observed as the pilot target for DUVET \citep{reichardt-chu_2022a} with KCWI \citep{kcwi} on Keck II on 15 February 2018 UT. The instrument was configured with the large slicer and two wavelength settings with the BM grating ($\lambda_c = 4,050$~{\AA} and $\lambda_c = 4,800$~{\AA}) to cover the full available wavelength range of KCWI ($3,500 \leq \lambda \leq 5,500$~{\AA}). This gave a field of view (FOV) of $33'' \times 20.4''$ and, with the default $2\times2$ on-chip binning, a spatial resolution of $0.29'' \times 1.35''$ and a spectral resolution of $R \sim 1,800$ (wavelength sampling of $0.5$~{\AA}~pix$^{-1}$). Two exposures of $600$~s each were obtained for the blue wavelength setting while four exposures of $100$~s, $300$~s, $600$~s, and $1,200$~s were obtained for the red wavelength setting to balance bright emission in the galaxy center and faint emission out to $7.4$~kpc from the galaxy. Separate sky frames were not obtained for this pointing and we did not use any dithering between exposures. 

We observed the CGM of IRAS08 on 18, 19, and 24 February 2020 UT with KCWI on Keck II. The instrument was configured with the large slicer and BL grating centered at $\lambda_c = 4,550$~{\AA}, resulting in a spectral resolution of $R \sim 900$ ($1$~{\AA}~pix$^{-1}$) covering $3,550 \leq \lambda \leq 5,580$~{\AA}. The field around IRAS08 was surveyed in four pointings with minimal overlap and angled such that the $33''$ side was oriented radially away from the galaxy to maximize the radial distance covered. Three of the pointings were co-located with the known filament of {\HI} gas in the direction of a nearby companion $60$~kpc away, while a fourth pointing was located at an azimuthal angle roughly $90^{\circ}$ to the east off the {\HI} filament (Fig.~\ref{fig:map}). The pointings have a radial extent from IRAS08 ranging $5.7-31$~kpc. Four exposures of $1,800$~s each were obtained for every pointing, with no dithering between exposures to minimize alignment errors since very few point sources are present in the FOV. With the expectation that emission from the CGM of IRAS08 would fill the FOV of each KCWI pointing, we required separate sky exposures to subtract the sky without removing signal. For this we obtained an additional $1,200$~s or $1,800$~s sky exposure each night located in an empty patch of sky beyond the $R_{\rm vir}$ of IRAS08.

\subsection*{KCWI Reduction}

To reduce the KCWI data, we used the IDL version of the KCWI Data Reduction Pipeline (DRP, v1.2.2; https://github.com/Keck-DataReductionPipelines/KcwiDRP) with standard settings for all exposures, however the sky subtraction methods differed between the galaxy and CGM pointings. The data were flux calibrated with standard stars Feige 66 (galaxy pointing), G193-74, or Feige 92 (CGM pointings). For the galaxy pointing, we used in-frame sky subtraction where emission lines and continuum from IRAS08 were masked since we did not obtain separate sky exposures during these observations. Emission is present in the galaxy pointing out to the edge of the FOV, which resulted in minor oversubtraction of the emission lines at the edges of the FOV. The oversubtracted flux ($\sim 5 \times 10^{-18}$~erg~s$^{-1}$~cm$^{-2}$ for {\OIII} and {\Hb}; ${\sim 5 \times 10^{-17}}$~erg~s$^{-1}$~cm$^{-2}$ for {\OII}) was later added back into these data by comparing the values to those in the nearest CGM pointing, which overlaps spatially with the galaxy pointing. 

For the CGM pointings, we skipped the DRP sky subtraction step to aid in further post-pipeline reduction. Specifically we used in-house software \citep{nielsen_2022} to correct the CGM pointings for a well-known $10\%$ wavelength-dependent illumination gradient across the KCWI FOV that is not fully removed by the DRP \citep{nielsen_2022, rupke_2019, cai_2019}. We make the assumption that the sky dominates the flux at nearly all wavelengths in the sky exposures and that it is constant across the FOV in every exposure. Because we only take one sky exposure for a given KCWI configuration on a given night, we also assume that the sky is unchanging for a set of sequential exposures. Any deviations from a flat background in the sky frames are then expected to be due to residual instrumental or reduction pipeline effects. This residual gradient primarily impacts background levels and the low surface brightness features we observe here when comparing one slice to another in a given exposure. To correct for this, the illumination gradient is estimated from the sky exposures and then corrected for in all exposures. The science exposures were not used for the gradient estimate because the FOV is aligned such that each subsequent slice is located at a larger radius from IRAS08. The galaxy has a strong continuum gradient in the nearest CGM pointing that decreases in the same slice-by-slice direction as the residual illumination gradient, which is difficult to disentangle from the instrumental gradient and would bias the correction. The sky exposures are located beyond the virial radius of the galaxy, where continuum emission is not present. We first masked objects identified in a sky exposure whitelight image to obtain only spaxels with a flux at the background level. We then measured the median smoothed spectrum in each of the $24$ slices in the masked sky cube separately, which were compared to an overall median smoothed spectrum for the masked sky cube to determine the illumination gradient variations slice-by-slice. This wavelength-dependent ratio measured from the sky exposures is divided out of both the sky and CGM pointing exposures slice-by-slice, resulting in cubes that are flat to less than $1\%$. 

The separate sky exposures for each CGM pointing were obtained during twilight, which slightly elevated the sky flux compared to the science exposures. While this does not impact the gradient correction since we are comparing slices to each other for a given exposure, it does impact the sky subtraction of science exposures obtained subsequently during the night. To adjust for this, we measured the median flux over $4,300 \leq \lambda \leq 4,700$~{\AA} (chosen to avoid emission lines) in masked sky and science exposures and scaled the sky by the ratio of these fluxes (roughly $0.91$ for the first two nights and $0.96$ for the third night). The median sky spectrum was then measured from the masked, flattened, and scaled sky cubes and subtracted from every spaxel in the corresponding CGM exposures.

\subsection*{Combining KCWI Exposures}

Given their differing observation strategies, the KCWI galaxy and CGM exposures were combined with slightly different methods. The four galaxy exposures with the red wavelength center ($\lambda_c = 4,800$~{\AA}) were optimally combined by replacing saturated spectral pixels in the longest exposure with the same pixels from the next longest exposure that was not saturated \citep{reichardt-chu_2022a}. This resulted in a cube where the $1,200$~s exposure dominates the spaxels beyond $r_{90}$ of the galaxy pointing, while the shorter exposures dominate the most central regions of the galaxy and the bright galaxy emission lines. We reprojected the resulting cube from rectangular to square spaxels ($0.29'' \times 0.29''$) with Montage \citep{montage} using a drizzle factor of $1.0$ and scaled the fluxes with the \verb+fluxScale+ option by the change in the spaxel dimensions to properly account for the redistribution of flux \citep{nielsen_2022}. The two exposures with the blue wavelength center ($\lambda_c = 4,050$~{\AA}) were reprojected and median combined with Montage using the same settings. Finally, the red and blue cubes were combined after scaling their fluxes to the mean flux of the two cubes in their overlapping spectral region ($4,357 \leq \lambda \leq 4,498$~{\AA}). We corrected the astrometry of this final cube to the HST/WFC3 F772W image using Source-Extractor \citep{source-extractor} to identify IRAS08 in both the HST image and a whitelight image from the combined KCWI cube. 

The CGM exposures were first corrected to the HST/WFC3 F772W image astrometry by comparing the coordinates of continuum sources in KCWI whitelight images obtained with Source-Extractor to those in the HST imaging. This corrected for slight shifts ($\sim 0.4''$) in object positions across the KCWI FOV between the first and last exposures for a given pointing. The exposures were then reprojected from rectangular to square spaxels and all four exposures for a given pointing were optimally combined with Montage using the same settings as for the galaxy pointing.

\subsection*{Emission line surface brightnesses}

To obtain accurate emission line fluxes and line ratios, we corrected all KCWI cubes for Galactic extinction \citep{cardelli_1989, schlafly_2011}. We also corrected for intrinsic dust extinction using the ratio between {\Hb} and {\Hg} \citep{calzetti_2001}, but only in the disk of IRAS08 where {\Hg} is detected. While dust is expected to be present in the CGM \citep{menard_2010}, the {\Hg} line, which is required to estimate the extinction levels, is no longer detected beyond $\sim 3$~kpc. Within $10-30$~kpc the average $V$-band extinction of the CGM in local Universe galaxies is estimated to be $A_V = 0.03$~mag \citep{menard_2010}. This would result in an extinction correction of only $\sim 3\%$ for the CGM spaxels, so dust extinction likely does not have a significant impact on the fluxes we measure in the CGM of IRAS08. 

The diffuse CGM’s low surface brightness results in low signal-to-noise, especially with increasing distance. To maximize the {\OII}, {\Hb}, and {\OIII} emission signal-to-noise ratio in the CGM, we first spatially binned all KCWI cubes $2 \times 2$ to account for the oversampling introduced by Montage, which also roughly matched the spaxel size to the seeing. We then Voronoi binned the cubes with VorBin \citep{vorbin} by enforcing a minimum signal-to-noise ratio of seven on the $\OIII~\lambda 5007$~{\AA} emission line. {\OIII} was chosen over {\OII} and {\Hb} as a compromise in that it is neither the brightest nor the faintest of the three lines within the disk, providing an optimal balance between signal-to-noise ratios and bin size while keeping the same bins for all three emission lines to allow for direct line ratio measurements. The signal-to-noise ratio in each KCWI spaxel was measured by first subtracting off the continuum around {\OIII} estimated with a third order polynomial fit across a bandwidth of $\lambda = 80$~{\AA}, where the line was masked within $\pm 7$~{\AA} of $z_{\rm gal}$. We summed the continuum-subtracted flux within this inner $14$~{\AA} region and the noise was estimated similarly from the variance cube within the same wavelength region. This line bandwidth was chosen to cover all of the emission line flux at the center of IRAS08, which has strongly blueshifted asymmetric emission lines due to starburst outflows \citep{reichardt-chu_2022a}, but to minimize the amount of noise included at larger radii. After binning, we remeasured the flux and noise of all three emission lines in every Voronoi bin with the same continuum and line bandwidths centered on $z_{\rm gal}$. We chose not to fit the emission lines with Gaussian profiles to avoid making an assumption on the line-of-sight kinematic structure of the CGM emission given the low spectral resolution. The fluxes were then converted to surface brightnesses by dividing by the area of the bin. The final cubes have average $3\sigma$ surface brightness limits on the continuum of $2 \times 10^{-18}$~erg~s$^{-1}$~cm$^{-2}$~arcsec$^{-2}$ in the galaxy pointing and $8 \times 10^{-19}$~erg~s$^{-1}$~cm$^{-2}$~arcsec$^{-2}$ in the CGM pointings. 

We quantified the change in the distribution of gas with radius in and around IRAS08 by fitting the surface brightness profiles in each line with two functions. The galaxy disk is well fit by an exponential function with form ${\rm SB} = a_{\rm exp}~e^{-r/r_{\rm 0,exp}}$, where $a_{\rm exp}$ is the surface brightness at the center of IRAS08 and $r_{\rm 0,exp}$ is the scale radius. The fitted parameters for each emission line are tabulated in Table~\ref{tab:fits}. IRAS08 has a scale radius of $\sim0.45$~kpc in all three emission lines. The CGM is instead well fit by a power law with the form ${\rm SB} = (r/r_{\rm 0,pow})^k$ out to at least $30$~kpc, where $r_{\rm 0,pow}$ is the radius at which the curve crosses a surface brightness of $10^{-16}$~erg~s$^{-1}$~cm$^{-2}$~arcsec$^{-2}$ and $k$ is the slope. The CGM slope is roughly $-2$ in {\OII} and $-1.5$ in {\Hb} and {\OIII}. {\OII} has the steepest decrease with radius while {\Hb} and {\OIII} have more shallow decreases. The profile changes from being dominated by an exponential in the galaxy disk to a power law in the CGM at radius $r_{\rm break}$, which represents the radius at which the two functions cross. This break radius is roughly $2r_{90}$ for every emission line.

\subsection*{Constraining the Electron Density with Mass}

The electron density is a key parameter for estimating the mass of the CGM and creating a full accounting of a galaxy's baryon budget \citep{werk_2014}. However, its value and spatial dependence are difficult to constrain. These properties dominate the uncertainty in deriving the mass of ionized CGM gas where, for example, an order of magnitude decrease in $n_e$ would result in an order of magnitude increase in mass. Our current observations of IRAS08 make directly measuring $n_e$ difficult, where the low spectral resolution of the KCWI data do not allow us to resolve the {\OII} doublet in the CGM. In the disk of IRAS08 the electron density can be estimated using the $\OII~\lambda\lambda 3727, 3729$ line ratio, which was found to be $n_e = 300-400$~cm$^{-3}$ \citep{reichardt-chu_2022b}. Outside the disk where we have sufficient spectral resolution, the {\OII} line ratio is below the low density limit \citep{kewley_2019}, which suggests $n_e < 30$~cm$^{-3}$, and CGM emission line stacking \citep{dutta_2023} also suggests a low $n_e < 100$~cm$^{-3}$. Methods for estimating $n_e$ with quasar absorption lines and fast radio bursts estimate $n_e\sim10^{-3}$~cm$^{-3}$ \citep{werk_2014, prochaska_2019}, which is several orders of magnitude smaller than the current emission estimates. 

To place limits on the emitting gas density in the CGM, we estimate the mass of the emitting CGM with the following equation:
\begin{equation}
    M_{\rm CGM} = \frac{1.36m_{\rm H}}{\gamma_{\Hbeq}~n_e} L_{\Hbeq},
\end{equation}
where $m_{\rm H}$ is the atomic mass of hydrogen. $\gamma_{\Hbeq}$ is the {\Hb} emissivity assuming that the gas has a temperature of $T \sim 10^4$~K and that case B recombination \citep{osterbrock_2006} applies ($1.24 \times 10^{-25}$~erg~cm$^3$~s$^{-1}$). This calculation depends on the gas {\Hb} luminosity, $L_{\Hbeq}$, which we can measure directly from our observations (Fig.~\ref{fig:profiles} top right), and electron density, $n_e$, which is currently unconstrained in the CGM. Both of these values could vary both radially from and azimuthally around the galaxy. The {\Hb} luminosity radial dependence is well constrained by the CGM power law, which we assume holds out to at least $R_{\rm vir}$. This assumption is reasonable since {\Ha} has been found to have a similar slope out to $1$~Mpc in stacked galaxies \citep{zhang_2016}. Azimuthal variations in {\Hb} luminosity are not supported by our observations since the surface brightness and ionization conditions are consistent regardless of whether the gas is co-located with the {\HI} filament (Figs.~\ref{fig:profile_zoom} and \ref{fig:profile_ratios}). We thus conservatively assume a sphere of gas with the power law radial dependence from Table~\ref{tab:fits} and azimuthal symmetry for {\Hb} luminosity. We also conservatively assume that $n_e$ has a constant radial and azimuthal value, exploring the range of masses obtained with values of $10^{-3} < n_e < 1$~cm$^{-3}$. We also assume a unity volume filling factor, although the gas is likely clumpy with both a radial and azimuthal dependence. 

These assumptions may break down to some degree, especially in the presence of outflows. Emission maps of edge-on galaxies show various minor axis outflow structures \citep{rupke_2019, cai_2019, burchett_2021, zabl_2021} and IRAS08 is known to have significant outflows directed perpendicular to its disk \citep{reichardt-chu_2022a, chisholm_2016}. Given the galaxy’s face-on inclination, we cannot constrain the outflow radial or azimuthal dependence. However, outflows are not expected to dominate the emission beyond the IRAS08 disk ($r_{\rm break}$) because (1) local starbursting galaxies typically have small opening angles ($\sim30^\circ$) \citep{shopbell_1998, westmoquette_2011, mcpherson_2023}, (2) the surface brightness of IRAS08's CGM is low compared to outflows in other galaxies \citep{rupke_2019, zabl_2021}, and (3) outflows tend to have $n_e \sim 50-700$~cm$^{-3}$ \citep{f-s_2020}, which is significantly larger than what we infer. Regardless, we are likely underestimating the CGM mass by neglecting the presence of outflows within $r_{\rm break}$.

\backmatter

\bmhead{Data Availability}

Raw Keck/KCWI data are publicly available at the Keck Observatory Archive (https://www2.keck.hawaii.edu/koa/public/koa.php) under program IDs W143, W185, and C232. Hubble Space Telescope imaging is publicly available on the Barbara A. Mikulski Archive for Space Telescopes under program ID GO-16749. Fully reduced data are available from the corresponding author upon request.

\bmhead{Acknowledgements}

We thank J.~X.~Prochaska for comments on the manuscript and the anonymous reviewers for constructive feedback that improved the manuscript. Parts of this research were supported by the Australian Research Council Centre of Excellence for All Sky Astrophysics in 3 Dimensions (ASTRO 3D), through project number CE170100013. R.R.V.~and K.S.~acknowledge funding support from National Science Foundation Award No.~1816462. Some of the data presented herein were obtained at the W.~M.~Keck Observatory, which is operated as a scientific partnership among the California Institute of Technology, the University of California and the National Aeronautics and Space Administration. The Observatory was made possible by the generous financial support of the W.~M.~Keck Foundation. Observations were supported by the joint Swinburne--Caltech Keck program 2020A\_C143 and Swinburne Keck programs 2018A\_W185, 2020A\_W143. The authors wish to recognise and acknowledge the very significant cultural role and reverence that the summit of Maunakea has always had within the indigenous Hawaiian community. We are most fortunate to have the opportunity to conduct observations from this mountain. This research is based on observations made with the NASA/ESA Hubble Space Telescope obtained from the Space Telescope Science Institute, which is operated by the Association of Universities for Research in Astronomy, Inc., under NASA contract NAS 5–26555. These observations are associated with program 16749. This research made use of Montage. It is funded by the National Science Foundation under Grant Number ACI-1440620, and was previously funded by the National Aeronautics and Space Administration's Earth Science Technology Office, Computation Technologies Project, under Cooperative Agreement Number NCC5-626 between NASA and the California Institute of Technology.

\bmhead{Author contributions}

N.M.N. and D.B.F. organized and wrote the main body of the manuscript. N.M.N. and G.G.K. led the observing proposal, observations, and planning of the galaxy KCWI observations, and D.B.F. provided the target selection. D.B.F. led the CGM KCWI observing proposal and N.M.N., G.G.K., and D.C.M. contributed to the proposal development and writing. N.M.N. and D.B.F. led the HST observing proposal and planning. N.M.N., D.B.F., G.G.K., D.C.M., B.R.C., K.M.S., and R.J.R.V. participated in the CGM KCWI observations and planning. D.C.M. built KCWI, developed the reduction pipeline, and provided technical expertise for the observations. N.M.N. developed the KCWI gradient removal tools, and performed the KCWI data reduction, surface brightness measurements, surface brightness profile fitting, and mass calculations. D.B.F. led the analysis and interpretation of the ionization conditions. G.G.K. and J.C. assisted in the interpretation of the results. B.R.C. developed code to perform the dust extinction corrections to the galaxy disk. All authors provided feedback to the manuscript.

\bmhead{Competing interests}

The authors declare no competing interests.



\newpage
\begin{appendices}

\section{Extended Data}\label{secA2}%

\renewcommand{\figurename}{Extended Data Figure} 
\begin{figure*}[h]
\centering
\includegraphics[width=\linewidth]{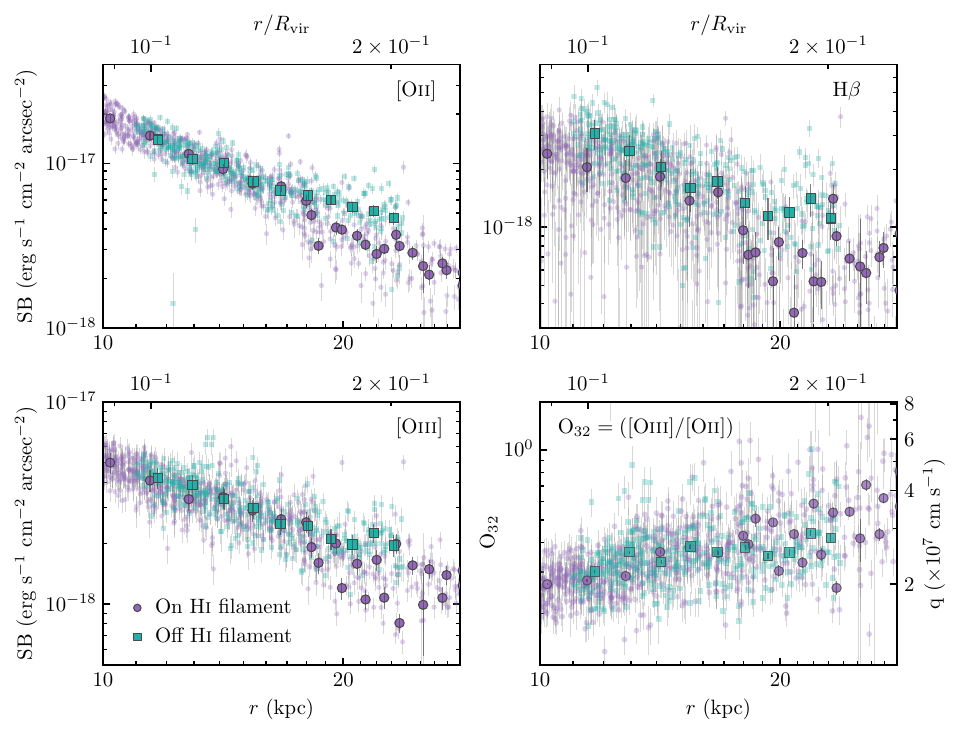}
\caption{{\bf Surface brightness and ionization do not vary significantly with azimuthal angle.} Points are colored as Fig.~\ref{fig:profiles}. The large points represent spaxels that have been binned radially for each KCWI pointing (10 radial bins per pointing) and correspond to the CGM points plotted in Fig.~\ref{fig:ratios}. Error bars on the binned data points are $1\sigma$ errors on the mean. Note that the y-axis ranges differ for each panel to emphasize the azimuthal variations and so the power law slope differences between lines are not reflected here.}
\label{fig:profile_zoom}
\end{figure*}

\begin{figure*}[h]%
\centering
\includegraphics[width=\linewidth]{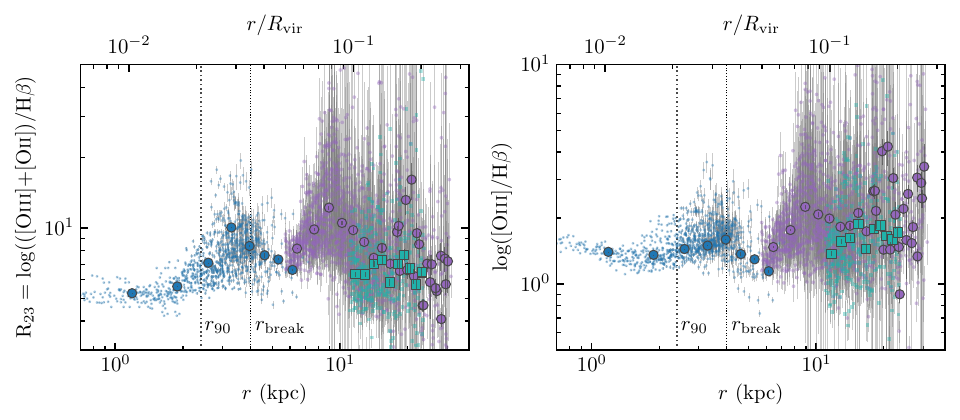}
\caption{{\bf Radial profiles for $\mathbf{{\rm R}_{23}}$ and {\OIII}/{\Hb}.} Points are colored as Fig.~\ref{fig:profiles}. The large points represent spaxels that have been binned radially for each KCWI pointing (10 radial bins per pointing) and correspond to the CGM points plotted in Fig.~\ref{fig:ratios}.}\label{fig:profile_ratios}
\end{figure*}

\end{appendices}



\end{document}